\documentclass[12pt,preprint]{article}
\usepackage{amsmath}
\usepackage{graphics}
\usepackage{epsfig}
\renewcommand{\vec}[1]{{\bf #1}}
\newcommand{\eqb}{\begin{equation}}
\newcommand{\eqe}{\end{equation}}
\newcommand{\dmb}{\begin{displaymath}}
\newcommand{\dme}{\end{displaymath}}
\newcommand{\pd}{\partial}

\newcommand{\eab}{\begin{eqnarray}}
\newcommand{\eae}{\end{eqnarray}}

\newcommand{\be}{\begin{equation}}
\newcommand{\ee}{\end{equation}}

\newcommand{\La}{\Lambda}
\begin{document}
\begin{titlepage}
\begin{flushright} 
\end{flushright}
\vspace{0.6cm}

\begin{center}
\Large{Yang-Mills thermodynamics: The deconfining phase}

\vspace{1.5cm}

\large{Ralf Hofmann}

\end{center}
\vspace{1.5cm}

\begin{center}
{\em Institut f\"ur Theoretische Physik\\ 
Universit\"at Frankfurt\\ 
Johann Wolfgang Goethe - Universit\"at\\ 
Robert-Mayer-Str. 10\\ 
60054 Frankfurt, Germany}
\end{center}
\vspace{1.5cm}

\begin{abstract}
We summarize recent nonperturbative results obtained for the 
thermodynamics of an SU(2) and an SU(3)
Yang-Mills theory being in its 
deconfining (electric) phase. Emphasis is 
put on an explanation of the concepts involved. 
The presentation of technical 
details is avoided.  

\end{abstract} 

\end{titlepage}

\noindent{\sl Introduction.} The purpose of a series of three articles, of which the
present one is the first, is to give a compact, abbreviated presentation of 
nonperturbative results obtained in 
\cite{Hofmann2005,HerbstHofmannRohrer2004,HerbstHofmann2004}. 
The objective of these papers is the 
thermodynamics of an SU(2) or an SU(3) Yang-Mills theory. 
Our goal is to provide the essential ideas and concepts avoiding 
technical detail. The discussion focuses on the 
case SU(2) for which we quote results 
explicitly. The steps needed to generalize to SU(3) and associated 
results are mentioned in passing only. The present paper discusses the 
deconfining or electric phase.

The key problem in addressing the quantum dynamics 
of an SU(N) Yang-Mills theory is to derive a mass 
gap in its spectrum, supposedly created by 
highly nontrivial ground-state dynamics. Although 
insightful in other circumstances diagrammatic approaches 
to this problem, which start out with a trivial ground state \cite{NP2004}, 
are probably hopeless since perturbation theory at best is an {\sl asymptotic} 
expansion in the coupling constant \cite{LargeOrderPT}. 

One-loop expressions for the quantum weight of isolated 
topological defects allow for the identification of certain nonperturbative 
mechanisms contributing to the ground-state physics 
\cite{SemiclassicalApprox}. 
Despite heroic efforts and partial successes a rigorous derivation 
of the Yang-Mills spectrum or of the complete spectrum of 
Quantum Chromodynamics seems to be out of reach within such a microscopic 
approach, see for example \cite{spectrum}. 

A more macroscopic but seemingly feasible approach to the 
problem of the mass gap is to consider a thermalized version of the 
theory, subjected to an appropriate spatial 
coarse-graining \cite{Hofmann2005,HerbstHofmannRohrer2004,HerbstHofmann2004}. 
One analyses the theory at high temperatures such that an evaluation of the 
partition function $Z$ in terms of the gauge fields appearing in the 
{\sl classical} action is suggestive. After a total 
gauge fixing one has 
\eqb
\label{partfunc}
Z=\sum_{A_\mu,\atop{A_\mu(\tau=0,\vec{x})=A_\mu(\tau=\beta,\vec{x})}}\,
\exp\left(-\frac{1}{2g^2}\int_0^\beta d\tau\int d^3x\,\,\mbox{tr}\,
F_{\rho\kappa}F_{\rho\kappa}[A_\mu]\right)
\eqe
where $\beta\equiv 1/T$, and $g$ is the gauge coupling. Instead of summing over all 
fluctuations in the field variables $A_\mu$ at once an approach 
roughly sketched in the following and made more explicit below is advantageous: 

First, eliminate the correlations mediated by noninteracting, 
classical saddle points of nontrivial 
topology (calorons and anticalorons) but trivial holonomy 
\cite{HarringtonShepard1977} in favor of a spatial average. (The term holonomy refers 
to the behavior of the Polyakov loop when evaluated on a 
gauge-field configuration at spatial infinity. A configuration is associated with a 
trivial holonomy if its Polyakov loop is in the center of the group ($Z_2$ for SU(2) and 
$Z_3$ for SU(3)) and associated with a nontrivial holonomy otherwise.) This 
is completely in line with Polyakov's early prophecy that certain classical 
solutions may cure the infrared catastrophe in Yang-Mills theory \cite{Polyakov1975}. 
(Recall, that in the limit 
$T\to 0$ a trivial-holonomy caloron approaches the BPST instanton in 
singular gauge \cite{BPST1976} and that the 
latter is interpreted as a large quantum 
fluctuation, that is, a topology-changing tunneling process. 
Recall also, that these processes can not 
be mimicked by a perturbative series 
due to the existence of an essential 
singularity in the associated 
weight at zero coupling. Since coarse-graining amounts to 
an average over {\sl all} quantum fluctuations it is imperative 
to include such large fluctuations in addition to 
the usual off-shell, plane-wave fluctuations considered in 
perturbation theory.) A spatially homogeneous, composite, and adjointly transforming 
scalar field $\phi$ emerges as a result of 
the applied spatial coarse-graining. This suggests the concept of a 
hypothetical energy- and pressure-free (BPS saturated)
thermal ground state composed of finite-action but 
energy- and pressure-free (BPS saturated) and 
noninteracting gauge-field configurations: 
trivial-holonomy calorons and anticalorons.  

Second, having derived an  
effective action for $\phi$ in a unique way establish the statistical and quantum mechanical 
inertness of this field: A spatial and partial moduli-space average over 
classical caloron and anticaloron configurations generates a 
classical (nonfluctuating) field $\phi$. 

Third, add the sector with 
trivial topology in a minimal fashion. In the emerging full effective theory 
quantum fluctuations, which deform the calorons and anticalorons and 
mediate interactions between them, 
are averaged over down to a spatial 
resolution of the order of the modulus $|\phi|$. 
Notice that $|\phi|^{-1}$ sets the linear size of the volume needed 
to perform the spatial average. Averaged-over quantum fluctuations 
with trivial topology occur in terms of a pure-gauge configuration $a_\mu^{bg}$ solving 
the effective Yang-Mills equations in the 
background of the field $\phi$. Together $\phi,a_\mu^{bg}$ comprise 
the full thermal ground state whose pressure $P^{gs}=-4\pi\Lambda_E^3 T$ is negative. On 
the microscopic level this can be understood as follows: Pairs of a BPS 
magnetic monopole and its antimonopole \cite{BPSmon} are contained within calorons or anticalorons of 
{\sl nontrivial} holonomy \cite{Nahm1984,LeeLu1998,KraanvanBaal1998}. The 
nontrivial holonomy is locally excited by gluon exchanges between 
calorons and anticalorons. A magnetic monopole and 
its antimonopole much more often attract 
than repulse each other as a result of quantum fluctuations 
around them \cite{Diakonov2004}. The spatial average over 
an attractive potential between a monopole and its antimonopole being 
in equilibrium with respect to their 
annihilation and recreation {\sl is} nothing but negative pressure. 

Fourth, compute the spectrum of 
propagating quasiparticle modes by virtue 
of the adjoint Higgs mechanism. The latter is the coarse-grained 
manifestation of off-Cartan gluons (unitary gauge) scattering 
off calorons and anticalorons along a zig-zag like path. An emerging, temperature dependent 
mass characterizes the propagation of such modes on distances 
larger than $|\phi|^{-1}$. This mass gap resolves the problem of 
the magnetic instability encountered in perturbation theory. Namely, the loop expansion of 
thermodynamical quantities converges very rapidly even though the gap approaches 
zero as $T^{-1/2}\propto |\phi|$ for $T\to\infty$. 
The rapid convergence is also due to residual quantum 
fluctuations being softer than $|\phi|$. (Only tiny corrections to the 
free-quasiparticle-gas approximation to thermodynamical quantities 
are induced by two-loop effects \cite{HerbstHofmannRohrer2004}.) 

Fifth, impose the requirement that the Legendre transformations between 
thermodynamical quantities are invariant under the applied 
spatial coarse-graining. This yields a first-order differential 
equation governing the evolution of the {\sl effective} gauge coupling 
with temperature. The identification 
of the phase boundary, where isolated monopoles become massless and thus 
condense, and the temperature dependence of thermodynamical quantities 
follow from this evolution. Let us now discuss 
each step in more detail.\vspace{-0.3cm}\\ 

\noindent{\sl The field $\phi$.} The color orientation 
$\hat{\phi}$ (phase) of the field $\phi$ is computed as a spatial and incomplete moduli-space 
average over the contributions of a trivial-holonomy 
caloron and a trivial-holonomy anticaloron in singular gauge \cite{HerbstHofmann2004}. 
While the definition of the associated 
dimensionless quantity 
\eab
\label{defphi}
\hat{\phi}^a\equiv\frac{\phi^a}{|\phi|}(\tau)\sim \int d^3x\,\int d\rho\,\,\frac{\lambda^a}{2}& F_{\mu\nu}[A_\alpha(\rho,\beta)]\left((\tau,0)\right)\,
\left\{(\tau,0),(\tau,\vec{x})\right\}[A_\alpha(\rho,\beta)]\times
\nonumber\\ 
&F_{\mu\nu}[A_\alpha(\rho,\beta)]\left((\tau,\vec{x})\right)\,
\left\{(\tau,\vec{x}),(\tau,0)\right\}[A_\alpha(\rho,\beta)]
\eae
is unique [no explicit $\beta$ dependence because the 
(anti)caloron action is $\beta$ independent; Wilson lines 
$\left\{(\tau,0),(\tau,\vec{x})\right\}[A_\alpha(\rho,\beta)]$, 
$\left\{(\tau,\vec{x}),(\tau,0))\right\}[A_\alpha(\rho,\beta)]$ are evaluated as 
functionals of the configuration $A_\alpha(\rho,\beta)$ 
along a straight path and no spatial shift $0\to \vec{z}$ because 
no associated mass scale available on the classical level; 
only trivial-holonomy calorons or anticalorons $A_\alpha(\rho,\beta)$ 
of topological charge modulus $|Q|=1$ are allowed 
because otherwise the larger number of dimensionful 
parameters in the moduli spaces would generate forbidden explicit $\beta$ dependences; 
no higher $n$-point contributions because they also would generate explicit $\beta$ dependences; 
flat measure for integration over scale parameter $\rho$ because the ratio of field and 
modulus is taken in Eq.\,(\ref{defphi})] a number of ambiguities 
emerge when evaluating the expression on the right-hand 
side in (\ref{defphi}): undetermined normalizations and shifts in the $\tau$ 
dependence of the caloron and anticaloron contribution and 
a global gauge ambiguity associated with the plane spanned by two unit vectors involved in 
an angular regularization \cite{HerbstHofmann2004} of the 
expression on the right-hand side of Eq.\,(\ref{defphi}). For a given angular-regularization 
plane (or, equivalently, for a given global gauge fixing) these ambiguities 
span the solution space of the differential equation ${\cal D} \phi=0$ where 
${\cal D}\equiv\pd_\tau^2+\left(\frac{2\pi}{\beta}\right)^2$. Thus the linear 
operator $\cal D$ and therefore $\phi$'s 
second-order equation of motion is {\sl uniquely} defined by the right-hand 
side in (\ref{defphi}). A requirement on $\phi$ is that its 
energy-momentum tensor vanishes in the effective theory 
because it emerges as an average over noninteracting field configurations 
whose (microscopic) energy-momentum tensor vanishes 
due to their (anti)selfduality. Since the $\tau$ 
dependence of the field $\phi=|\phi|\hat{\phi}$ resides in 
its phase $\hat{\phi}$ and since $\cal D$ is a {\sl linear} operator 
this is equivalent to requiring $\hat{\phi}$'s BPS saturation. Thus we have 
\eqb\label{1orderbps}
\pd_\tau \hat{\phi}=\pm\frac{2\pi i}{\beta}\lambda_3\,\hat{\phi}\,
\eqe
where $\lambda_3$ is the third Pauli matrix. Because of the afore-mentioned 
global gauge ambiguity $\lambda_3$ can be replaced by any 
normalized linear combination of Pauli matrices. The solutions to Eq.\,(\ref{1orderbps}) 
are given as 
\eqb\label{sol1orderbps}
\hat{\phi} = C \, \lambda_1 \, 
\exp\left(\mp \frac{2\pi i}{\beta} \lambda_3 (\tau -\tau_0)\right)\,.
\eqe
Evidently, 
the number of ambiguities has reduced from four to two after requiring 
BPS saturation for the trajectory within a given angular regularization plane. 
The parameter $\tau_0$ can globally be gauged to zero. However, the normalization $C$ is 
undetermined on the classical level. (Recall, 
that we have not yet considered interactions between 
calorons and anticalorons.) This is a consequence of the unbroken 
symmetry of the classical Yang-Mills action 
under spatial dilations. 

After plane-wave quantum 
fluctuations are integrated out down to a resolution $|\phi|$ this symmetry is broken by 
dimensional transmutation \cite{NP2004}. To determine the ($\tau$ independent) 
modulus $|\phi|$ we assume that the corresponding mass scale $\Lambda_E$ 
is externally provided such that calorons and anticalorons remain free of 
interactions. (Together, $\beta$ and $\Lambda_E$ set the spatial length 
scale $|\phi|^{-1}$ over which the coarse-graining involving a single 
caloron and a single anticaloron of topological charge modulus $|Q|=1$ ought to be carried out.) 
Then the entire field $\phi$ satisfies a BPS equation with a right-hand 
side whose 'square' uniquely defines the potential 
appearing in the effective action: Since the latter is an average 
over noninteracting calorons down to a resolution set by the modulus $|\phi|$ itself 
the potential (and its square root) 
ought not depend on $\beta$ explicitly. To reproduce the 
known $\tau$ dependence of the phase $\hat\phi$ the right-hand side of 
$\phi$'s BPS equation must be linear in $\phi$. 
At the same time it should be an analytic 
function of $\phi$. Then, up to global gauge rotations, 
the only viable possibility is \cite{HerbstHofmann2004}
\eqb\label{bps14}
\pd_\tau\phi=\pm i\,\Lambda_E^3\,\lambda_3\,\phi^{-1}\,
\eqe
where $\phi^{-1}\equiv \frac{\phi}{|\phi|^2}$. 
Substituting $\phi=|\phi|\hat{\phi}$ into 
Eq.\,(\ref{bps14}) with $\hat{\phi}$ given in Eq.\,(\ref{sol1orderbps}), 
we have $|\phi|(\beta,\Lambda)=\sqrt{\frac{\beta\Lambda_E^3}{2\pi}}$. 'Squaring' 
the right-hand side of Eq.\,(\ref{bps14}) one 
obtains the unique potential $V_E=\mbox{tr}\frac{\La_E^6}{\phi^2}$. (Shifts $V_E\to V_E+\mbox{const}$ 
are forbidden by Eq.\,(\ref{bps14})!) Because at a given temperature $T$ the mass squared 
$\pd^2_{|\phi|}\,V_E(|\phi|)$ of possible fluctuations $\delta\phi$ is much 
larger than the resolution squared $|\phi|^2$ and also 
than $T^2$ one concludes that $\phi$ is an inert background in the effective 
theory \cite{Hofmann2005}: Quantum and statistical fluctuations $\delta\phi$ are absent! Notice 
that both Eq.\,(\ref{sol1orderbps}) and 
Eq.\,(\ref{bps14}) are not invariant under 
Lorentz boosts but invariant under spatial rotations and spacetime shifts. 
This is a manifestation of the fact 
that the isotropic and spacetime homogeneous heat bath defines a 
preferred rest frame. \vspace{-0.3cm}\\ 

\noindent{\sl The full ground state.} 
After a minimal extension of $\phi$'s action involving the coarse-grained sector 
with trivial topology one derives the associated equation of motion 
\eqb\label{eomts}
D_\mu G_{\mu\nu}=ie[\phi,D_\nu \phi]
\eqe
where $e$ is the {\sl effective} gauge coupling. 
Notice that no gauge-field configuration with localized action 
density exists as a solution to the classical field equations 
in the effective theory due to the 
spatial homogeneity of the field $\phi$. In particular, 
no topologically nontrivial configuration of finite action exists. 
The pure-gauge configuration $a_\mu^{bg}=\frac{\pi}{e}\,T \delta_{\mu 4}\,\lambda_3$ 
solves  Eq.\,(\ref{eomts}) by virtue of $D_\mu \phi=0=G_{\mu\nu}$: A remarkable 
thing has happened. Coarse-grained interactions, mediated by 
off-shell gluons mediating a resolution large than $|\phi|$, emerge 
as a pure-gauge configuration $a_\mu^{bg}$. The latter induces a finite ground-state 
energy density $\rho^{gs}$ and a finite ground-state pressure 
$P^{gs}$ as $\rho^{gs}=-P^{gs}=4\pi\Lambda_E^3 T$. This makes the 
so-far hidden scale $\Lambda_E$ (gravitationally) detectable. A 
microscopic interpretation of this result is 
available owing to the heroic work in \cite{Diakonov2004}.   
 
The one-loop quantum weight for an SU(2) caloron or anticaloron  
with nontrivial holonomy, topological charge modulus $|Q|=1$, and 
no net magnetic charge \cite{Nahm1984,LeeLu1998,KraanvanBaal1998} was computed 
in \cite{Diakonov2004}. 
Depending on the holonomy is was shown that the magnetic 
monopole-antimonopole constituents either attract (small holonomy) or repulse (large holonomy) 
each other implying that a semiclassical argument excluding these (instable) 
saddle points from the partition function 
by a vanishing quantum weight in the 
thermodynamical limit is flawed \cite{GPY1981}. Since the 
small-holonomy solution collapses 
back to trivial holonomy (attracting and annihilating monopole and antimonopole) 
its quantum weight essentially coincides with the one for zero holonomy. The latter, however, 
is finite in the thermodynamical limit. The likelihood for exciting a large 
holonomy, which facilitates a life in isolation for 
monopoles and antimonopoles by quantum induced repulsion \cite{Diakonov2004} and subsequent 
screening \cite{KorthalsAltes}, is roughly given by $e^{-40}\ll 1$ \cite{Hofmann2005}. So the local 
creation of a small holonomy by gluon exchanges between trivial-holonomy 
calorons is the by-far dominating situation: At high temperatures the 
ground state of the Yang-Mills theory is characterized 
by an equilibrium with respect to annihilation and recreation of attracting 
monopoles and antimonopoles, and a negative ground-state 
pressure emerges from spatially averaged attraction between monopoles and antimonopoles. 

An important observation concerning the deconfining nature of the electric phase 
is the $Z_2$ degeneracy of the coarse-grained Polyakov loop $\cal P$. 
This can be seen by identifying the singular, periodic transformation to unitary gauge, 
which maps ${\cal P}[a_\mu^{bg}=\frac{\pi}{e}\,T \delta_{\mu 4}\,\lambda_3]=-{\bf 1}$ 
to ${\cal P}[a_\mu^{bg}=0]={\bf 1}$ on tree-level, by showing the invariance of the periodicity of 
a gauge-field fluctuation $\delta a_\mu$ under this transformation, 
and by recalling that the field $\phi$ breaks the gauge symmetry 
SU(2) to U(1) after coarse-graining. An immediate consequence of the latter fact is that 
there are massless modes in the effective theory assuring the qualitative validity of 
the tree-level result for $\cal P$ also on the 
level of the full expectation. For SU(3) the field $\phi$ 
winds in each SU(2) subalgebra for one third of the 
extent $\beta$ of the time interval. Along the same lines as in the SU(2) case 
the $Z_3$ degeneracy of the Polyakov-loop expectation follows after an identification of 
the irreducible representation of this group in 
terms of possible tree-level values of $\cal P$.\vspace{-0.3cm}\\ 

\noindent{\sl Almost free quasiparticles.} 
Propagating, topologically 
trivial excitations $\delta a_\mu$ are affected 
by the ground state in two ways. 

First, a part of the 
spectrum acquires mass by the adjoint Higgs mechanism (thermal quasiparticle modes). For 
SU(2) two directions in color space (off-Cartan in unitary gauge) 
acquire the same mass $m=2e|\phi|$. For SU(3), 
where coarse-graining breaks the gauge symmetry SU(3) to U(1)$^2$, the common 
mass $m_2=2e|\phi|$ emerges for two color directions, and four directions acquire 
the same mass $m_4=e|\phi|$ \cite{Hofmann2005}. 
It is known for a long time that the nonconvergence of the perturbation series for infrared 
sensitive thermodynamical quantities arises from the 
weakly screened magnetic sector, that is, 
from the absence of an infrared cutoff \cite{Linde1980}. 
This problem is resolved in the effective theory. 

Second, residual plane-wave quantum fluctuations only probe 
distances larger than $|\phi|^{-1}$. In a physical gauge 
(unitary-Coulomb) this imposes a cutoff 
on the maximal off-shellness for quantum fluctuations in the gauge field (ultraviolet cutoff). 

To summarize, the spatial coarse-graining over 
{\sl all} quantum fluctuations causes the simultaneous emergence of 
tree-level masses (infrared cutoffs) 
and constraints on the maximal off-shellness 
of plane-wave quantum fluctuations (ultraviolet cutoff). 
In the deconfining phase this renders the excitations of an SU(2) and of an SU(3) 
Yang-Mills theory to form a gas of very weakly interacting quasiparticles 
\cite{HerbstHofmannRohrer2004}. \vspace{-0.3cm}\\ 

\noindent{\sl Evolution of the effective coupling.} 
There are two undetermined parameters in the effective theory, the Yang-Mills 
scale $\Lambda_E$ and the effective gauge coupling $e$. While the former represents 
a constant mass scale, which embodies the coarse-grained manifestation 
of dimensional transmutation, the latter may well depend 
on temperature. The temperature evolution of $e$ is governed by 
the invariance of Legendre transformations between thermodynamical quantities 
under the spatial coarse-graining applied. One has \cite{GorensteinYang1995}
\eqb\label{tsc}
\pd_m P=0
\eqe
where $m$ is 
the mass gap in the quasiparticle spectrum, and $P$ denotes the total 
pressure (excitations plus ground state). For many practical purposes it suffices 
to use the one-loop expression for $P$. Eq.\,(\ref{tsc}) implies a first-order 
evolution equation for temperature as a function of mass whose solution 
can be inverted to extract the dependence of the effective 
gauge coupling $e$ on temperature. 
\begin{figure}
\begin{center}
\leavevmode
\leavevmode
\vspace{5.5cm}
\includegraphics{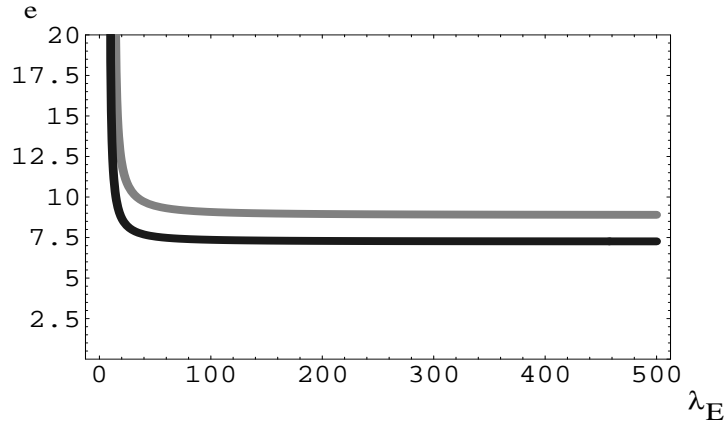}
\end{center}
\caption{The temperature evolution of the gauge 
coupling $e$ in the electric phase for 
SU(2) (grey line) and SU(3) (black line). The gauge coupling 
diverges logarithmically, $e\propto -\log(\lambda_{E}-\lambda_{c,E})$, at 
$\lambda_{c,E}=13.867$ (SU(2)) and $\lambda_{c,E}=9.475$ SU(3) 
where $\lambda_{E}\equiv \frac{2\pi T}{\Lambda_E}$. 
The respective plateau values are $e=8.89$ and $e=7.26$.\label{Fig1}}      
\end{figure}
In Fig.\,\ref{Fig1} this dependence is 
depicted for SU(2) and SU(3). It is a striking result that the downward evolution at low 
temperatures, where $e$ approaches a plateau and subsequently runs into a 
logarithmic divergence, $e\sim -\log(T-T_{c,E})$, is {\sl independent} 
of the boundary condition imposed at a high temperature. This is the celebrated 
ultraviolet-infrared decoupling property which so far was only shown 
in perturbation theory. Naively seen, the plateau values $e=8.89$ (SU(2)) 
and $e=7.26$ (SU(3)) render the effective 
theory strongly coupled and the one-loop analysis useless. This conclusion is 
incorrect, however, since interactions are mediated by severely constrained 
gluon exchanges. Thus spatially coarse-grained gluons interact only very weakly and 
one-loop expressions are reliable on the 0.1\%-level \cite{HerbstHofmannRohrer2004}. 
The constancy of the coupling $e$ for temperatures well 
above $T_{c,E}$ on average expresses the conservation of the magnetic charge 
of isolated, screened, and nonrelativistic 
monopoles ($g=\frac{4\pi}{e}$ for a single monopole) 
inside the spatial volume of typical size $|\phi|^{-3}$, and the behavior 
$e\sim -\log(T-T_{c,E})$ signals the vicinity of a phase 
transition where coarse-grained off-Cartan modes decouple and 
isolated magnetic monopoles become massless and thus condense.\vspace{-0.3cm}\\ 

\begin{figure}
\begin{center}
\leavevmode
\leavevmode
\vspace{5.5cm}
\includegraphics{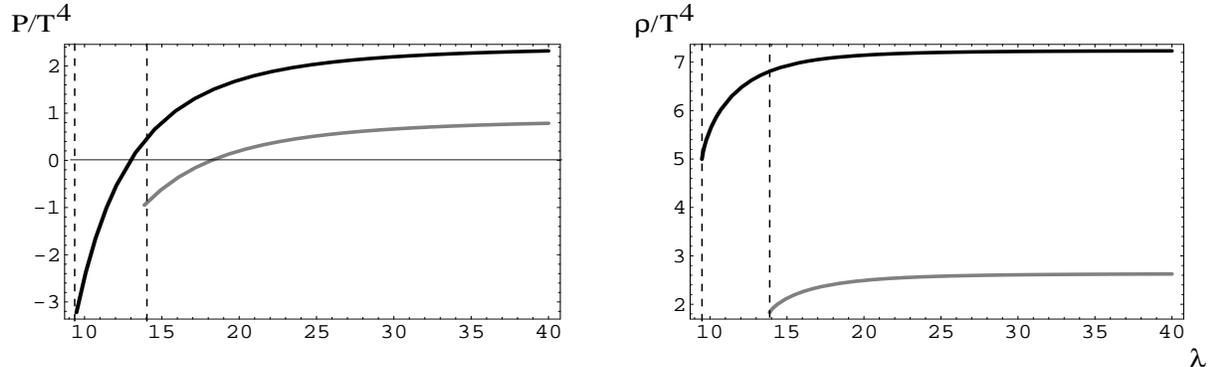}
\end{center}
\caption{The ratios of energy density $\rho$ and pressure 
$P$ with $T^4$ for SU(2) (grey lines) and 
SU(3) (black lines). The vertical, dashed lines indicate 
the critical temperature for the onset of 
magnetic monopole condensation.\label{Fig-2}}      
\end{figure}

\noindent{\sl Results and summary.} 
In Fig.\,\ref{Fig-2} we present results for the temperature 
evolution of the total pressure $P$ and the total 
energy density $\rho$. 
Notice the rapid (power-like) approach 
to the asymptotic regime, which up to a small mismatch in 
the number of polarizations, coincides with the naive Stefan-Boltzmann 
limit (gas of massless gluons with two polarizations each). Notice also the 
dip in $\rho$ at $\lambda_{c,E}$. (Towards 
the magnetic side of the phase boundary the energy density jumps 
up due to an increase of the number of polarizations 
for the propagating gauge modes: A coupling to averaged-over, 
condensed magnetic monopoles renders these modes massive 
and breaks the dual gauge symmetry U(1)$_D$ (SU(2)) and U(1)$^2_D$ (SU(3)) 
down to its discrete subgroup.) If the U(1)$_Y$ factor in the 
Standard Model is an effective manifestation of SU(2) 
dynamics subject to a Yang-Mills scale $\Lambda_{\tiny\mbox{CMB}}$, which is 
comparable to the temperature $T_{\tiny\mbox{CMB}}$ of the cosmic microwave background, then 
this dip is experimentally determined to be 
at $T_{\tiny\mbox{CMB}}=\frac{\lambda_{c,E}}{2\pi}\Lambda_{\tiny\mbox{CMB}}$ \cite{Hofmann2005}. 
(Only at $T_{\tiny\mbox{CMB}}$ is the photon entirely 
unscreened and are (stable) off-Cartan modes entirely decoupled). 
The dip dynamically stabilizes $T_{\tiny\mbox{CMB}}$ 
against (accelerated) cosmological expansion dominantly induced by 
another matter source (slowly 
rolling Planck-scale axion \cite{Hofmann2005,Wilczek2004}).

\section*{Acknowledgements}
The author would like to thank Mark Wise for useful conversations 
and for the warm hospitality extended to him during his 
visit to Caltech in May 2005. Financial support by 
Kavli Institute at Santa Barbara and by 
the physics department of UCLA is thankfully acknowledged.

\baselineskip25pt
\end{document}